# Pathway-Activity Likelihood Analysis and Metabolite Annotation for Untargeted Metabolomics using Probabilistic Modeling

**Ramtin Hosseini, Neda Hassanpour, Li-Ping Liu♦, and Soha Hassoun♦,*** 

Tufts University; Department of Computer Science, Medford, MA, 02155
{Ramtin.Hosseini, Neda.Hassanpour, Liping.Liu, Soha.Hassoun}@tufts.edu

♦ Equal Contributions

* Correspondence: Soha.Hassoun@cs.tufts.edu

**Abstract**

**Motivation:** Untargeted metabolomics comprehensively characterizes small molecules and elucidates activities of biochemical pathways within a biological sample. Despite computational advances, interpreting collected measurements and determining their biological role remains a challenge.

**Results:** To interpret measurements, we present an inference-based approach, termed Probabilistic modeling for Untargeted Metabolomics Analysis (PUMA). Our approach captures metabolomics measurements and the biological network for the biological sample under study in a generative model and uses stochastic sampling to compute posterior probability distributions. PUMA defines a pathway as *active* if the likelihood that the path generated the observed measurements is above a particular (user-defined) threshold. PUMA predicts the likelihood of pathways being active, and then derives a probabilistic annotation, which assigns chemical identities to the measurements. PUMA is validated on synthetic datasets. When applied to test cases, the resulting pathway activities are biologically meaningful and distinctly different from those obtained using statistical pathway enrichment techniques. Annotation results are in agreement to those obtained using other tools that utilize additional information in the form of spectral signatures. Importantly, PUMA annotates many additional measurements.

**Availability:** The code and datasets are on https://github.com/HassounLab/PUMA

## 1. Introduction

Analyzing cellular responses to perturbations such as drug treatments and genetic modifications promises to elucidate cellular metabolism, leading to improved outcomes in personalized medicine and synthetic biology. Metabolomics has emerged as the new 'omics', providing a read out of cellular activity that is most predictive of phenotype. Metabolomics so far have played a critical role in advancing applications spanning biomarker discovery [1], drug discovery and development, [2], plant biology [3], nutrition [4] and environmental health [5]. Importantly, the advent of *untargeted metabolomics* to measure molecular masses and spectral signatures of thousands of small molecule metabolites for a biological sample allows unprecedented opportunities to characterize the phenotype.

The success of untargeted metabolomics in providing insight into cellular behavior however hinges on solving two problems. *Metabolite annotation*, concerns associating measured masses with their chemical identities. This problem is challenging, as a particular mass may be associated with multiple chemical formulas (*e.g.*, there are 21,988 known molecular formulas associated with $C_{20}H_{24}N_2O_3$). There are several techniques for annotating measurements. Database lookups rely on comparing the measured spectral signature against experimentally generated fragmentation patterns cataloged in reference spectral databases (*e.g.*, METLIN [6], HMDB [7], MassBank [8], NIST [9]). Database coverage however is limited as catalogued spectral signatures are obtained experimentally. Alternatively, computational methods that either mimic the ionization and fragmentation process or utilize machine learning techniques (*e.g.*, MetFrag [10], Fragment Identificator (FiD) [11], CFM-ID [12], CSI:FingerID [13]) score the measured spectra against the predicted spectra of molecules in a candidate set. The chemical identity associated with the highest scoring signature(s) is then assigned to the measured spectra. Other annotation techniques exploit the biological context of the measurements. iMet[14] and BioCAN [15] exploit data about local neighborhoods within the network graphs to improve annotation.

The second problem, *pathway enrichment analysis*, concerns interpreting measurements within their biological context to study coordinated changes arising in response to cellular perturbations. Overrepresentation Analysis (ORA) tools (*e.g.,* MESA [16], MetaboAnalyst [17], MPEA [18]) employ statistical testing (*e.g.,* Fisher's exact test) to determine if a pathway is enriched in measured metabolites to a degree different than expected by chance when compared to other cellular pathways or those in a reference sample [19]. Pathway enrichment techniques can be broadly classified in two categories. Topological Analysis (TA) compute the observed metabolites' centrality and connectivity, metrics that reflect the importance of a metabolite in the turnover of molecules through a pathway or network (e.g., MetaboAnalyst [17] and IMPaLA [20]). Metabolite annotation and pathway enrichment have traditionally been solved as two independent problems, where pathway enrichment assumes that the chemical identity of each measured mass is known *a priori*. In general, pathway analysis techniques therefore do not adequately address issues related to uncertainty in metabolite annotation. One exception is Mummichog, a set of statistical algorithms that predict functional activity directly from measurements considered significant when compared to those in a reference sample [21].

We present a novel inference-based probabilistic approach, Probabilistic modeling for Untargeted Metabolomics Analysis (PUMA), for interpreting metabolomics measurements. One input to PUMA is the set of measurements that are already processed through metabolomics data processing workflow, e.g., MZmine [22] or XCMS [23]. Another input is a set of pathways, each consisting of enzymatic reactions and their metabolic products, that are specific to the sample under study. Such pathways can be readily assembled from databases such as KEGG or MetaCyc or others. Using these data, PUMA first predicts the likelihood of activity of metabolic pathways within a biological sample using. PUMA then utilizes these predictions to derive probabilistic assignment of measurements to candidate chemical identities. PUMA utilizes inference and approximates posteriors using Gibbs Sampling, a Monte Carlo Markov Chain (MCMC) sampling technique [24]. Although inference is a well-known machine learning technique, there were several challenges in developing PUMA including: 1) identifying a suitable generative model that represents the underlying biological process, 2) expressing complex relationships using probability distributions, 3) speeding the inference procedure with complex mathematical marginalization and vectorization, 4) identifying best model parameters, and 5) validating model against the ground truth. Herein, we describe how PUMA addresses such challenges. PUMA is then applied to two data sets collected for Chinese Hamster Ovary (CHO) cells [15] and human urinary samples [25]. Predicted pathway activities are analyzed for biological significance and compared against activity predictions obtained through statistical pathway enrichment analysis. For the CHO cell test case, metabolite annotations obtained using PUMA are compared to those obtained using methods that utilize the spectral signature to annotate metabolites (HMDB [7], METLIN, [6] and BioCAN [15]). For the human urinary samples, PUMA annotations are compared to published annotations obtained using spectral databases and experimental validation.

## 2. Methods

### 2.1 Motivating Example

A small example is provided to illustrate challenges in mapping measurements to metabolites and pathways, and to show inference's ability to address these issues. **Figure** 1 presents a snippet of a network that shows two pathways (ovals), Pathway 1 and Pathway 2. Metabolites with known chemical identities associated (circles) are either associated with one pathway (red circle) or more than one pathway (blue circles). Measurements (squares) correspond to masses that can be associated with one particular metabolite (red square) or multiple metabolites (blue squares). Not all metabolites within a sample are measured due to either instrument limitations or because they are simply not present in the sample due to biological or environmental factors. Some metabolites are thus not associated with any measurements (white circles), and maybe associated with one or more pathways.

There are two types of uncertainties in interpreting measurements from untargeted metabolomics. One type of uncertainty relates to assignment of metabolites to pathways (circles to ovals, **Figure** 1). For example, measurement $w_3$ is assigned to metabolite $j_5$. Because $j_5$ is a metabolite common to both Pathways 1 and 2, there is an uncertainty in assignment of the metabolite to the pathways: $j_5$ can be the product of activity in either Pathway 1 or Pathway 2. The other uncertainty relates to assignment of masses to metabolites, when a mass can map to multiple metabolites (squares to circles, **Figure** 1). Measurement $w_4$ can be attributed to one or two metabolites, $j_6$ and $j_7$, both sharing the same mass. The uncertainty in assigning $w_4$ to metabolites $j_6$ and $j_7$ manifests in further uncertainty. If $w_4$ is associated with $j_6$, then it contributes to the activity of Pathway 1 (and/or other pathways with which $j_6$ is associated), while, if $w_4$ is associated with $j_7$, then it contributes to the activity of Pathways 2 (and/or other pathways with which $j_7$ is associated). Not all measurements contribute to these uncertainties. For example, measurement w₅ is unique to metabolite j₁₃. In turn, j₁₃ is unique to Pathway 2. Some measurements (such as w₅) clearly contribute more significantly than others (such as w₃ and w₄) in determining pathway activities.

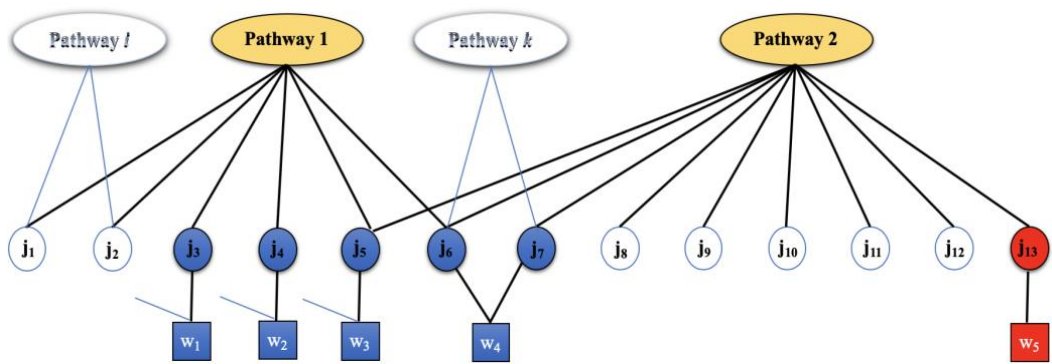

**Figure 1. Illustrative example of uncertainty when mapping measurements to metabolites and pathways.** Pathways (ovals) are associated with metabolites (circles), which in turn are associated with measurements (square). White circles represent non-measured metabolite with membership in one or more pathways. Blue circles represent measured metabolites that have multiple-pathway memberships (multiple-pathway membership is assumed but not shown for j₃ and j₄). The red circle represents a metabolite that has membership in only one pathway. Measurement w₅ uniquely maps to j₁₃, which uniquely maps to Pathway 2, while all other measurements map to multiple metabolites, as shown by solid or dotted lines.

Computing pathway activities using an enrichment ratio can be misleading, because it does not take into account the uncertainty in attributing measurements to metabolites and pathways. The enrichment ratio for Pathway 1 can be computed as the ratio of 4 putatively measured metabolites divided by 6 total metabolites in the pathway. While this enrichment ratio seems high, there is little confidence that Pathway 1 is active since all measured metabolites form this pathway could be due to active pathways other than Pathway 1. Pathway 2 has an enrichment ratio equal to 3 divided by 8. The significance or importance of this ratio is unclear. Inference will conclude that Pathway 2 is active with high probability, as it includes a measured metabolite that cannot be attributed to the activity of any other pathway. In contrast to enrichment methods, our inference-based technique considers uncertainties in measurement-metabolite and metabolite-pathway relationships when computing the likelihood of pathway activities. A pathway is considered *active*, if the likelihood that it generated the observed measurements is above a particular threshold. When we analyze our test cases, we will assume a threshold of 0.5. A user of PUMA may decide to use this threshold or select a more suitable threshold above which pathways are deemed active.

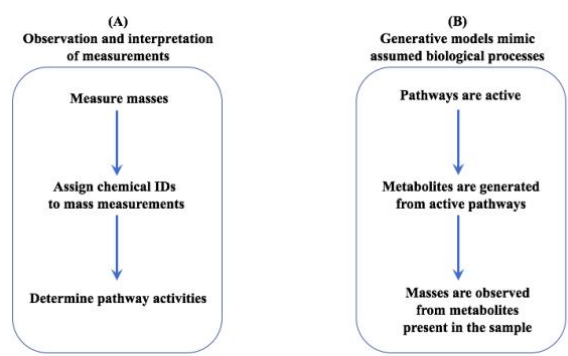

**Figure 2.** Comparison of a workflow to collect and interpret observations (A), and a generative model that captures a biological process (B).

*2.2 Generative Model*

To determine pathway activities, an untargeted metabolomics workflow (**Figure** 2A) begins with collecting measurements, followed by metabolite annotation using annotation tools (e.g. database look ups or annotation tools) and then applying pathway analysis tools (e.g. ORA or TA) to determine pathway activities. A pathway is assumed active when biological and environmental factors lead to the production of some or all of its metabolic products. In some cases, metabolite annotation is skipped, and statistical pathway activity is computed directly from measurements. In contrast, our inference-based approach utilizes a generative model (**Figure** 2B) that mimics biological processes inherent to the sample under study. Our presumed biological process assumes that when pathways are active, they cause the presence of some its metabolites, which in turn results in observations of masses through untargeted metabolomics.

PUMA first constructs a graphical model [26] that captures the complex relation among pathway activities, metabolites, and measurements in a single integrated model. The model produces values that are observed (measured), as well as hidden variables of interest, which cannot be directly observed but rather inferred from those values that can be observed. In our case, the observations correspond to mass measurements collected through untargeted metabolomics. The hidden variables are pathway activities and the presence of a metabolite in a biological sample.

Our generative model assumes the following biological process: one or more pathways are active. An active pathway causes the presence of some of its metabolites, which in turn results in observations of masses through untargeted metabolomics data collection. The generative model is parameterized with prior information, or prior probabilities, about the behavior of the biological process. Here, we provide priors on each step in the biological process: for pathway activities, on pathways generating their metabolites, and metabolites mapping to mass measurements. We assume that the biological sample has a metabolic model with $I$ pathways, $J$ metabolites and $K$ unique metabolite masses. A metabolite may have membership in one or more pathways. A measured mass may be associated with one or more masses of the model metabolites. Metabolite masses are discretized by $K$ bins. Each bin is centered at a unique mass value and allows for a mass tolerance of +/-15 ppm. Each metabolite is assigned to a single bin that is centered closest to the metabolite's mass. A binary vector $w$ has $K$ entries and indicates mass observations of metabolites in the model. A 1 entry for $w_k$ in vector $w$ indicates the observation (measurement) of at least one metabolite in the $k$th bin while a 0 indicates no observation for any metabolite in that bin.

Let $a = (a_i : i = 1, \dots, I)$ denote the status of $I$ pathways in the biological sample, so $a$ is a vector of binary random variables, where a value of 1 indicates that the corresponding pathway is active and 0 indicates inactivity. We assume that the $a_i$ random variables are independent, with a Bernoulli($\lambda$) prior:

$$p(a_i = 1) = \lambda, \qquad i = 1, \dots, I \tag{1}$$

For simplicity in defining our model, we assume that $\lambda$ is a model parameter and set it to a constant. As an alternative, we can give it a Beta prior.

Matrix $O$ is defined with $I$ rows and $J$ columns. Each entry $o_{ij}$ corresponds to the activeness of metabolite $j$ in pathway $i$, where a value of 1 indicates metabolite $j$ is active due to pathway $i$ and a value of 0 indicates that metabolite $j$ is not produced by pathway $i$. If a metabolite $j$ is on a pathway $i$, then the metabolite is produced according to the following probability.

$$p(o_{ij} = 1 | a_i = 1) = \mu, \qquad p(o_{ij} = 1 | a_i = 0) = 0 \tag{2}$$

Otherwise, $p(o_{ij} = 1 | a_i) = 0$ when $j$ is not on $i$. For simplicity, we assume that all metabolites are equally likely to be generated with probability $\mu$ within an active pathway. Vector $m$ collapses the matrix $O$ into a binary vector with $J$ elements, indicating the activeness of a metabolite due to whichever pathway.

$$m_j = \left[ \sum_i o_{ij} > 0 \right] \tag{3}$$

Here $[\cdot]$ gives 1 when the condition inside is true or 0 otherwise.

As not all masses can be captured using the mass spectrometer, its observed accuracy is defined using parameter $\gamma$. Let $J_k$ define the group of metabolites that have masses in the $k$-th bin, then

$$p(w_k = 0 | m_{J_k}) = (1 - \gamma)^{\sum_{j \in J_k} m_j} \tag{4}$$

This probability means that every metabolite present in the biological sample has a chance $\gamma$ to be detected. In the case when all metabolites in $J_k$ are not observed ($\sum_{j \in J_k} m_j = 0$), then mass $k$ will not be observed. The detection of a metabolite is independent of the detection of others in the sample. No two groups, $J_k$ and $J_{k'}$, intersects because a metabolite has only one mass. The described model is described using the plate representation [27] (**Figure 3**). The model presents the joint probability distribution of random variables $a$, $o$, $m$ and $w$ defined as:

$$p(a, o, m, w) = p(a; \lambda) \, p(o|a; \mu) \, p(m|o) \, p(w|m; \gamma) \tag{5}$$

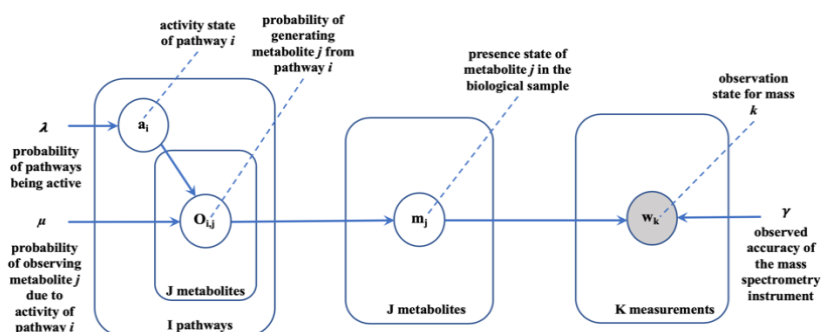

**Figure 3. Graphical representation of the generative model.** To avoid representing all *I* pathways, *J* metabolites and *K* masses in the graph, we use the 'plate' notation and draw one representative node per variable and enclosing these variables in a plate (rectangular box). The number of instances of each enclosed variable is indicated by the fixed constant in the lower right corner of the box. Random variables of the model *(a, o, m, w)* are shown in white circles. The variable *m* has a deterministic relationship with *O*. The shaded circle, labelled *w*, represents an observed random variable. $\mu, \lambda, \gamma$ are parameters to the model.

## *2.3 Inference*

Using the probabilistic model, we infer pathway activities and metabolite presence from mass measurements. Specifically, we calculate the following probabilities. For each pathway $i$ in the biological sample we calculate $p(a_i|w)$, the posterior probability of pathway $i$ being active given evidence in mass measurements. PUMA utilizes Gibbs sampling to perform Bayesian inference [24] to approximate the posterior probabilities of pathway activities conditioned on the measurements. We then infer the presence of metabolites by calculating the posterior $p(m_j|w)$ for all $j$. We use the latter probabilities to rank a candidate set of metabolites for each mass measurement, where a candidate set provides one or more suggestion of chemical identities that have the same mass, within an error margin, as the observed one.

### 2.3.1 Inferring pathway activities

Gibbs sampling is employed to perform Bayesian inference to approximate $p(a|w)$, the posterior probability of pathway activities conditioned on the measurements. Naively sampling random variables $a$ and $O$, is time consuming. To speed the Gibbs sampler, we marginalize hidden variables $O$. From the Bayesian formula,

$$p(a|w) = p(w|a)p(a)/p(w) \qquad (6)$$

Gibbs sampling is convenient in that it there is no need to compute the denominator $p(w)$ to draw samples from the posterior $p(a|w)$. We only need to focus the computation of $p(w|a)$ and $p(a)$, where the latter was already assumed to have a Bernoulli distribution. Below we show how to compute $p(w|a)$. We point out that $p(w|a)$ decomposes as follows:

$$p(w|a) = \prod_k p(w_k|a) \qquad (7)$$

This is because metabolites in separate $J_k$ groups are independent given $a$, so do masses that are computed within these groups. Then we focus on the calculation of $p(w_k|a)$. Let $\phi_j(a)$ be the probability that at least one pathway in the biological sample generates metabolite $m_j$. That is, $\phi_j(a) = p(m_j = 1|a)$. The detailed calculation of $\phi_j(a)$ is provided in the **Supplementary File 1**, the calculation of $\phi_j(a)$ is:

$$\phi_j(a) = 1 - (1-\mu)^{n_j} \qquad (8)$$

with $n_j$ being the number of active pathways that $j$ is on. Probability $p(w_k|a)$ is then computed as follows:

$$p(w_k|a) = \begin{cases} 1 - \prod_{j \in J_k}[1 - \gamma\phi_j] & w_k = 1 \\ \prod_{j \in J_k}[1 - \gamma\phi_j] & w_k = 0 \end{cases} \qquad (9)$$

The expression $1 - \gamma\phi_j$, a number between 0 and 1, represents the likelihood that the mass spectrometer did not measure the activity of metabolite $m_j$. Combining $p(w_k|a)$ with the Bernoulli prior $p(a)$, we have the joint probability $p(w, a)$, which is sufficient for running the sampler and getting samples from the posterior. If $\lambda$ has a Beta prior, then we will sample $a$ and $\lambda$ together from $p(\lambda)p(a|\lambda)p(w|a)$.

### 2.3.2 Inferring metabolite annotations

With samples drawn from $p(a|w)$, we approximate $p(m_j|w)$, the posterior probability distribution of metabolite $j$ being present in the biological sample. Instead of running the Gibbs sampling procedure again, we use previously collected samples of $a$ from $p(a|w)$ to estimate the probability $p(m_j|w)$. Let $S = \{\, a \in samples\ of\ p(a|w)\}$ be a set of samples from the distribution $p(a|w)$, then

$$p(m_j|w) = \Sigma_a p(m_j, a|w) = \Sigma_a p(m_j|a, w) p(a|w) \approx \frac{1}{|S|} \Sigma_{a \in S} p(m_j|a, w) \tag{10}$$

The probability $p(m_j|a, w)$ has efficient computation. Let $k_j$ denote the entry of $w$ corresponding to metabolite $j$, and let $\setminus k_j$ denote other entries in $w$. Then:

$$p(m_j|a, w) = \frac{p(m_j, w|a)}{p(w|a)} = \frac{p(m_j, w_{k_j}|a)\, p(w_{\setminus k_j}|a)}{p(w|a)} \tag{11}$$

Here we use the fact that $m_j$ and $w_{k_j}$ are independent of other mass observations when $a$ is given. With this relation, we have:

$$p(m_j = 1|a, w) = \frac{p(m_j = 1, w_{k_j}|a)}{p(m_j = 1, w_{k_j}|a) + p(m_j = 0, w_{k_j}|a)} \tag{12}$$

Here the terms that are constants to $m_j$ are canceled. Finally, we can compute $p(m_j, w_{k_j}|a)$ by marginalizing over all $m_{j'}$ for $j' \neq j$ and $j' \in J_k$:

$$p(m_j,\ w_{k_j}|a) = \Sigma_{m_{J_k \setminus j}} p(m_j, m_{J_k \setminus j}, w_{k_j}|a) = \Sigma_{m_{J_k \setminus j}} p(w_{k_j}|m_j, m_{J_k \setminus j}) p(m_j, m_{J_k \setminus j}|a) \tag{13}$$

We decompose the above formulation into two terms for managing calculations. These two terms, $p(w_k|m_{J_k})$ and $p(m_j, m_{J_k \setminus j}|a)$ are further derived and re-expressed in the supplementary material, to yield the following:

$$p(m_j,\ w_{k_j}|a) = \begin{cases} (1 - \phi_j)\left(\prod_{j' \in J_k,\ j' \neq j}(1 - \gamma \phi_{j'})\right) & m_j = 0,\ w_{k_j} = 0 \\[6pt] (1 - \phi_j)\left(1 - \prod_{j' \in J_k,\ j' \neq j}(1 - \gamma \phi_{j'})\right) & m_j = 0,\ w_{k_j} = 1 \\[6pt] \phi_j(1 - \gamma)\prod_{j' \in J_k,\ j' \neq j}(1 - \gamma \phi_{j'}) & m_j = 1,\ w_{k_j} = 0 \\[6pt] \phi_j\left(1 - (1 - \gamma)\prod_{j' \in J_k,\ j' \neq j}(1 - \gamma \phi_{j'})\right) & m_j = 1,\ w_{k_j} = 1 \end{cases} \tag{14}$$

We use these equations to calculate the probabilities $p(m_j = 1,\ w_{k_j}|a)$ and $p(m_j = 0,\ w_{k_j}|a)$. By normalizing the two terms to have a sum of 1, we get the posterior of metabolite annotations. The derived probabilities are used as a scoring metric to rank a candidate set for each mass measurement. Details on the derivation and implementation of metabolite annotation are provided in **Supplementary File 1**.

*2.4 Implementation and parameter initialization*

We implemented PUMA using PyMC3 [28], a probabilistic programming framework that allows for automatic Bayesian inference on user-defined models. In the implementation, we assume that $\lambda$ has a $\beta$ prior with parameters $\alpha = \beta = 1$. We sample both random variables $a$ and $\lambda$. To draw samples from a posterior distribution, PyMC3 utilizes a Markov Chain Monte Carlo (MCMC) sampling technique [29]. The generative model was derived from the metabolic model for each of our case studies. The observed accuracy of the mass spec, $\gamma$, is assumed to be 0.9. Each entry in $\mu$ is assumed to be 0.5 if metabolite $j$ exists on pathway $i$. $T$, the number of samples to draw from the model, is a variable that can be set in PyMC3. The sampler was run multiple times with $T$ values equal to 500, 1000 and 1500. For all reported runs, increasing the number of drawn samples did not affect the computed probabilities for pathways activities. Results are reported for sample sizes of 1000.

**3. Results**

*3.1 Case study 1: Chinese Hamster Ovary (CHO) Cell*

We apply PUMA to LC-MS (liquid-chromatography mass spectrometry metabolomics data for CHO cell cultures belonging to a low growth cell line [15] (**Supplementary Table S1**). This dataset was well annotated using BioCAN[15], a tool that aggregates results from spectra databases and annotation tools. The CHO cell data was collected separately under three different combinations of liquid chromatography methods and positive or negative ionization modes. When combined, the data provides a more comprehensive characterization of the sample in the form of 8,711 measurements. The metabolic model for the CHO cell was culled from KEGG [30], based on unique metabolites and pathways for the cricetulus griseus (Chinese hamster) under organism code *cge*. The model has 86 pathways, 1,534 metabolites, and 722 unique mass measurements. The model has 86 pathways, 1,534 metabolites, and 722 unique mass measurements. Due to incompleteness of metabolic models, there were only 635 metabolites that map to 411 mass measurements in the combined dataset. The observed masses for the model are used to initialize the observation vector $w$ for each dataset.

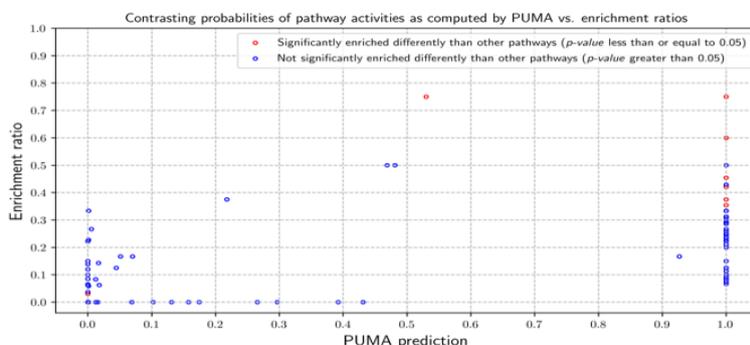

**Figure 4. probability of pathway activities as computed by PUMA vs. enrichment ratios for CHO cell**. Each data point is marked as either statistically enriched (red) or non-statistically enriched (blue) based on a Fisher's Exact Test *p*-values of 0.05.

### 3.1.1 Probabilities of pathway activities

Detailed results for each dataset and for the combined data set is provided in **Supplementary Table S1**. A pathway is considered active if $p(a_i|w)$ is equal to or greater than 0.5. As mass observations differ from one set of measurements to another, the predicted activity differs among the datasets. A detailed discussion of the results for the individual datasets is provided in **Supplementary File S1**. The rest of the CHO cell analysis provided here is based on the combined dataset.

Many of the 42 pathways identified active by PUMA are biologically relevant. The biological activity of most pathways such as TCA cycle, essential for energy metabolism, Biotin (vitamin B7) metabolism, amino acid synthesis, and many others, is expected. However, the activity of some pathways including caffeine and drug pathways is biologically unlikely active in the CHO cell samples. Based on our experiments using the synthetic datasets, we expect some PUMA predictions to be false.

Pathway activities predicted by PUMA are contrasted against pathway enrichment ratios (**Figure 4**). The *enrichment ratio* for a particular pathway is defined as the ratio of measured masses that map to metabolites within the pathway to its size. Pathways are labeled as *statistically enriched* based on statistical significance of their ratios using Fisher's Exact Test (FET). The null hypothesis is that there is no difference between the enrichment ratios of pathways in the sample. A *p-value* equal to or less than 0.05 is considered significant. Eight pathways are designated statistically enriched. These pathways are Galactose metabolism, Fatty acid degradation, Purine metabolism, N-Glycan biosynthesis, Amino sugar and nucleotide sugar metabolism, Glycosaminoglycan degradation, Glycerophospholipid metabolism, lipoic acid metabolism. Among them, 6 pathways were predicted by PUMA to be active with probability equal to 1 while the N-Glycan biosynthesis pathway had a 0.53 likelihood of being active. Fatty acid degradation is predicted to be inactive. There were many pathways that had low enrichment ratios and low PUMA-predicted activity.

While there was consensus in some cases, there were also differences. PUMA designates some pathways as active despite low enrichment ratios. For example, the enrichment ratios of the TCA cycle, fatty acid biosynthesis, ubiquinone and terpenoid-quinone biosynthesis are 0.15, 0.29, and 0.13, respectively. Meanwhile, PUMA predicted these pathways active with a likelihood of 1. There are three pathways with enrichment ratio equal to 0.5. Of them, one pathway, biotin metabolism, is assigned active by PUMA with probability 1.0. The biotin metabolism pathway has a measured mass that is unique and cannot be generated by other pathways. However, the other two pathways, both glycosphingolipid biosynthesis pathways, are predicted active with probability less than 0.5 (0.47 and 0.48). The reason was as follows: the observed mass measurements in the glycosphingolipid biosynthesis pathways could be mapped to Galactose metabolism and Glycosaminoglycan degradation pathways that are associated with a unique measurement that cannot be attributed to any other pathway in model (similar to the case of $w_5$ in our illustrative example **Figure 1**). As the result, the glycosphingolipid biosynthesis pathways were assigned probabilities less than 0.5, while the pathways with the unique measurements are predicted active with high probability.

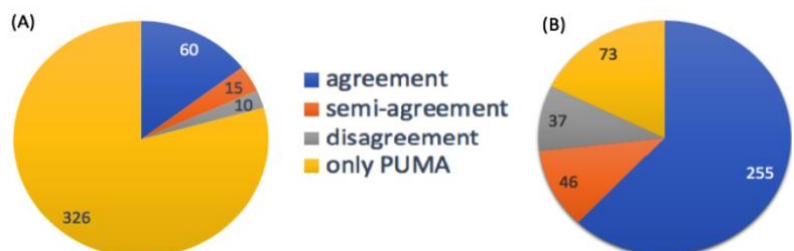

**Figure 5.** Metabolite annotations attained with PUMA against those identified by: **(A)** searching spectral databases, HMDB and METLIN, and **(B)** BioCAN. The blue slice in each pie represents "agreement". The orange and gray slices represent "semi-agreement" and "disagreement" respectively. Finally, the yellow slice represents the number of mass measurements that could only be annotated by PUMA.

### 3.1.2 Probabilities of metabolite annotations

A particular measurement was associated with a model metabolite if its mass matched the measured mass within the bin tolerance. Each measurement therefore may be assigned zero, one or more possible annotation. Probabilities of each metabolite being present in the sample as inferred by PUMA are used to score and rank the putative annotations. Here, only the top ranked metabolite(s) for each mass is considered as the *PUMA candidate set.*

We assess the accuracy of PUMA annotations by comparing the level of agreement of PUMA annotations with those using two other techniques, spectral database searches and BioCAN (**Figure 5**). Spectral signatures collected through untargeted metabolomics were looked up in METLIN and HMDB, and were previously reported [15]. The highest scoring metabolites for each measurement in METLIN and in HMDB formed the spectral database candidate set. Out of 411 mass measurements, 85 were identified as either in HMDB or METLIN. For each measurement, the PUMA candidate set was compared against the candidate set identified by HMDB and METLIN. The comparison leads to four different scenarios. One scenario is "agreement", where the PUMA candidate set exactly matches the candidate set from HMDB and METLIN. Such agreement occurs in 60 cases. There are 15 cases of "semi-agreement", where the candidate set from HMDB and METLIN is a subset of the top candidate set obtained from PUMA annotation. Three are 10 cases of "disagreement", where the candidate set from METLIN and HMDB does not overlap with the PUMA candidate set. In 7 such cases, the candidate metabolite from METLIN and HMDB is the second likely putative annotation identified by PUMA. These putative annotations, which were not included in the PUMA candidate set, had a high activity score and close to that of the metabolite(s) in the candidate set. In the remaining three cases, however, the candidate metabolite from METLIN and HMDB is assigned a low score by inference-based annotation workflow, a score far from the one assigned to the metabolite in the PUMA candidate set. These three cases are considered as genuine disagreement in annotation. Importantly, in the final scenario, "Only PUMA", with 326 cases, there were no matching annotations in METLIN and HMDB, reflecting the low coverage of spectral databases.

PUMA annotations are compared against those obtained using BioCAN [15]. BioCAN aggregates results from spectral database searches and *in silico* fragmentation tools and estimates the confidence in an annotation for a mass measurement not only based on a consensus but also by the confidence of presence of metabolites that are connected to the mass measurement through substrate-product relationships. BioCAN annotates 338 out of 411 mass measurements that are annotated by PUMA. We analyze the various scenarios as we did when comparing against spectral database annotations. There are 255 cases of agreement, 46 cases of semi-agreement, 37 cases of disagreement, and 73 new annotations by PUMA. The disagreements fell into two categories. In 17 out of 37 cases, there was disagreement on the top candidate, where PUMA ranked BioCAN's candidate as second best. There were genuine disagreements in 20 cases were the annotation by BioCAN was assigned a low score by PUMA.

In summary, comparing PUMA annotations against those obtained through spectral database and BioCAN shows significant levels of agreement. METLIN, HMDB and BioCAN incorporate spectra signatures during annotation while PUMA relies solely on pathway organization and mass measurements. Importantly, for the CHO cell, PUMA increased annotation by 383% over spectral databases and by 21% over BioCAN.

### 3.1.3 Evaluation of PUMA in overcoming uncertainty in annotation

Our synthetic dataset analysis indicated robustness to the uncertainty inherent in mapping measurements to metabolites when analyzing pathway activities using inference. The experiment is repeated using the annotation data for the CHO cell from METLIN and HMDB. For each mass $k$ annotated using METLIN or HMDB as metabolite $j$, matrix $\tau$ is modified. Column entries other than $\tau_{i,j}$ are set to zero, indicating that mass $k$ uniquely maps to metabolite $j$. Using the updated $\tau$, PUMA calculated posteriors for pathway activities. There was a slight change in predicted posteriors (average increase of 0.003) compared to those obtained using the original $\tau$ matrix. The change however does not alter posterior probabilities sufficiently to modify the list of active pathways. We repeated the analysis but incorporated the annotation data available from BioCAN instead of that obtained through spectral databases. The change in $\tau$ caused a slight change in predicted posteriors (an average of 0.001 per pathway) compared to those obtained using the original $\tau$ matrix. The one significant change was for pathway Phenylalanine metabolism where pathway activity changed from 0.03 to 1.0. The Phenylalanine metabolism pathway is responsible for producing Tyrosine. this finding shows that substantial additional annotations, as provided in the form of added annotations by BioCAN over the use of spectral databases, are required to inform inference in regard to pathway activities. Importantly, the results are in agreement with those for the synthetic dataset: annotating metabolites first has limited impact on the accuracy of computing pathway activities.

*3.2 Case study 2: human urinary sample*

We apply PUMA to untargeted metabolomics datasets collected for human urinary samples analyzed by Roux et al. [25]. Detailed annotations are provided for 384 measurements. The metabolic model for the urinary sample was derived from BioCyc [31]. The model had 275 pathways, 716 metabolites, and 565 unique masses. Only 123 out of the Roux et al. measured masses matched to those in the model.

3.2.1 Probabilities of pathway activities

PUMA designated 41 pathways as active in human urinary sample (**Supplementary Table S1**). We investigate how inference results compare with pathway enrichment ratios (**Figure 6**). Of the 41 pathways designated to be active using PUMA, six pathways (tRNA charging, 4-hydroxyproline degradation I, histidine degradation VI, lysine degradation II, purine ribonucleosides degradation to ribose-1-phosphate, nicotine degradation III) are statistically enriched. As in the CHO cell cases, there were cases of agreement and disagreement. There are several pathways were PUMA predicts low activity, while enrichment assumes a high enrichment ratio, including alanine biosynthesis II, glutamate degradation II, aspartate biosynthesis, arginine degradation VI and alanine degradation III. The probabilities for these pathways are 0.26, 0.22, 0.17, 0.31 and 0.25, respectively, while the corresponding enrichment ratios are 1.0, 0.57, 0.75, 0.6 and 1.0. Many measurements assigned to these pathways, however, are not unique as they can generated due to activity of other pathways.

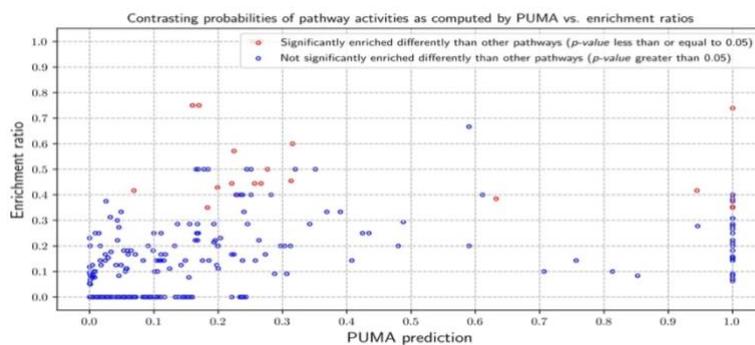

**Figure 6. probability of pathway activities as computed by PUMA vs. enrichment ratios for the human urine sample.** Each data point is marked as either statistically enriched (red) or non-statistically enriched (blue) based on a Fisher's Exact Test *p*-values of 0.05.

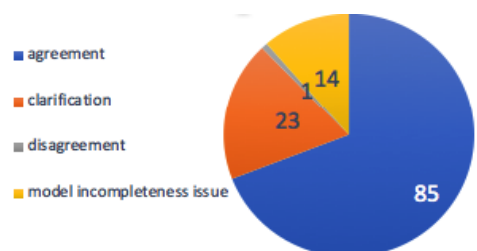

**Figure 7. Metabolite annotations attained with PUMA against those identified by Roux et al.** The blue slice represents "agreement". The orange slice represents "clarification". The gray slice represents "disagreement" and the yellow slice represents "model incompleteness issue".

3.2.2 Probabilities of metabolite annotations

The PUMA probabilities for each metabolite being present in the sample are used to score and rank metabolites. Only the top ranked metabolite(s) for each mass are considered as the PUMA candidate set. We compared our annotation against those provided by Roux et al. [25] (**Figure 7**). These annotations were either identified by matching at least two of their physicochemical parameters to those in a reference standard or annotated through spectral database lookups (HMDB). Some measurements were annotated as isomers, without identifying the precise chemical molecular identity. Of the 108 measured masses that matched to metabolites in the model, there were 85 cases of "agreement", where PUMA predictions matched the Roux et al. annotations. There were 23 cases of "clarification", where PUMA provided a specific chemical annotation while Roux et al annotated the measurement as an isomer. There was one case of "disagreement", where a Roux et al. annotation was predicted not present by PUMA. Finally, there were 14 "model incompleteness issue" cases where Roux et al assigned the measurement a chemical identity that was not in the model, indicating that PUMA provides the best match within the scope of model metabolites. We expect that more comprehensive metabolic model could address such issues.

*3.3 Model validation*

To give confidence in the performance of PUMA, it is desirable to validate the generative models against a "ground truth" dataset, where all measured metabolites are annotated and there is sufficient experimental evidence to allow attributing measured metabolites to specific pathways. Predictions by PUMA can then be compared against this ground truth. Despite several databases that catalogue various metabolomics datasets (e.g., Metabolights [32] and Metabolomics Workbench [33]), there are currently no untargeted metabolomics sets that are 100% annotated. Further, there are no datasets that allow attributing metabolites to specific pathways through experimental work. In lieu of such unavailable "ground truth" datasets, we generated synthetic metabolomics datasets from presumed known biological processes to validate our generative models. As central metabolism and network topology is conserved across many organisms [34], we generated the synthetic datasets for a representative organism, the CHO cell, discussed in the prior case study. CHO cell is a popular organism utilized in many biological studies.

Several synthetic datasets were generated. A random portion (0.3, 0.5, and 0.7) of pathways are assumed active, and a random portion (0.05, 0.10, 0.15, 0.20, 0.25, 0.50, 0.75) of metabolites within each active pathway are generated. For each portion of active pathways and for each portion of active metabolites, 100 metabolomics datasets reflecting the masses of the active metabolites were generated. The observed accuracy $\gamma$ was set to 1. We applied PUMA to each dataset and averaged PUMA's precision, recall and accuracy on identifying the presumed active pathways. At a pathway activity of 0.3 (**Figure S1A**), as we have more observed metabolites, recall increases because PUMA has more evidence in terms of observations to recover the correct pathway activities. Precision, PUMA's ability to label true positives correctly, is greater than 0.71, regardless of the active fraction of metabolites. Accuracy improves with increased active metabolites due to the corresponding increase in PUMAs ability to identify true positives. This trend holds for other assumptions about pathway activities (**Figures S2A and S3A**).

We investigate how uncertainty in metabolite annotation impacts inference regarding pathway activity. Before running PUMA, each mass measurement is attributed to a presumed active metabolite, thus removing annotation uncertainty. Results (**Figures S1B, S2B, and S3B**) show a similar trend to those in **Figures S3A, S4A, and S5A**. A similar trend holds when each measured mas is randomly assigned a metabolite amongst model metabolites with the same mass as a measured mass (**Figures S1C, S2C, and S3C**). This result emphasizes that computing pathway activities without the explicit step of performing metabolite annotation via spectral databases or annotation tools is a profitable approach. PUMA can therefore be used to accelerate the process of pathway activity analysis by direct use of mass measurements and bypassing metabolite annotation using spectral databases.

We further investigated the robustness of the model to its parameters. While prior runs assumed that the probability of observing a metabolite due to a particular pathway activity was 0.5, we varied the corresponding model parameter $\mu$ to 0.25 and to 0.75 and re-ran PUMA. The results (**Figure S4**) show that inference is dominated by other aspects of the model and that inference is robust to this model parameter.

*3.4 Model convergence, complexity and runtimes*

Drawing 1000 samples was used as a default. To ensure conversion of the Gibbs sampler, the number of samples was doubled until changes within the results were less than 0.01. The time and space complexity in sampling the model is $O(T \times I \times J)$. The runtime for drawing 1000 samples for pathway activity prediction and metabolite annotation for the CHO cell dataset were 231 and 0.5 seconds, respectively. The corresponding runtimes for the Human Urinary case study were 280 and 0.4 seconds, respectively. The runs were performed on a Dell PowerEdge R815 server with 64 cores (4x AMD Opteron 6380 processors) and 128Gb of RAM, running at 2.5GHz.

**4. Discussion**

We presented in this paper PUMA, a probabilistic approach to interpret mass measurements collected through untargeted metabolomics. PUMA fist uses inference to determine pathway activities. While prior works focused on computing pathway enrichment in the context of comparing one sample against the other, here, we define a pathway as active based on its likelihood of being responsible for the presence of one or more metabolomics measurements. In determining activity, PUMA reasons about the complex relationships between the measurements as well as known pathway as defined through the underlying biochemical networks. In doing so, levels of uncertainty in mapping measurements to metabolites and pathways are significantly reduced. Moreover, a clearer view of the likelihood of pathway activity levels emerges when compared to simple enrichment analysis. PUMA then utilizes the likelihood of pathway activities to compute the posterior probability distribution of metabolites being present in the sample.

The approach of predicting functional activity directly from spectral features without a priori metabolite annotation was previously shown effective for Mummichog [21]. PUMA utilizes all measurements to compute the likelihood that pathways gave rise to the measurements for a biological sample under a certain condition, while Mummichog utilizes differentially expressed metabolites to determine differently observed pathways. PUMA uses the likelihood of pathway activities to derive metabolite annotations. PUMA confirms that the organization of metabolic networks can resolve the ambiguity in metabolite annotation to a large extent, as previously illustrated in Mummichog.

PUMA is based on inference but differs from other inference-based methods. ProbMetab [35] uses a probabilistic method [36] to assign empirical formulas to measured spectra given potential formulas. The method proposed by Jeong et al constructs a generative model to infer the likelihood of a metabolite in the sample and the correctness of matching the measurement to a candidate metabolite within a spectral database based on measured spectra's similarity to that of the proposed candidate and to other competing spectra in the database [37]. The competing spectra, however, may not be relevant to the sample. Del Carratore et al uses evidence in the form of isotope patterns, adduct relationship and biochemical connections to infer metabolite annotations [38]. ZODIAC [39] also utilizes inference to re-rank molecular formula candidates suggested by SIRIUS [40].

PUMA was applied to two case studies, the CHO test case and the human urine test case. In both cases, PUMA offered a perspective on pathway activity that is distinctly different from that offered by statistical enrichment. PUMA identifies pathways that have a high likelihood of being active but have statistically low enrichment ratios, and pathways with low activity probabilities yet with statistically high enrichment ratios. Importantly, because inference reduces the uncertainty in mapping measurements to chemical identities, PUMA was able to successfully improve annotation. For the CHO cell tests case, PUMA was able to infer pathway activity levels similar to those identified with additional annotation information from other tools. Further, PUMA results had high agreement to annotations using spectral database lookups and BioCAN. This high level of agreement occurs despite the fact that PUMA does not utilize additional information in form of spectra signatures, as employed other techniques. In the case of the CHO cell test case, PUMA increased the percentage of mass annotation by 383% over spectral lookups and by 21% over BioCAN. For the human urine test case, PUMA showed agreement in annotating 85 metabolites that were prior annotated using database looks ups. PUMA also suggested 23 new identities that were previously identified as isomers. Importantly, the agreements shown in both test cases against prior experimental annotations (those provided by BioCAN [15] and by Roux et al [25]) allowed us to validate the utility of PUMA.

**Supplementary Materials:** See supplementary online material for additional information on derivations in support of inference of pathway activity inference and metabolite annotations, and some results on the synthetic dataset and on the CHO case study.

**Funding:** Research reported in this publication was supported by the National Institute of General Medical Sciences of the National Institutes of Health under Award Number R01GM132391. The content is solely the responsibility of the authors and does not necessarily represent the official views of the National Institutes of Health.

**Acknowledgments:** We are grateful to Alex Tong and Jan-Willem van de Meent who suggested several preliminary ideas on using inference to determine pathway activity. We are also grateful to Nicholas Alden for providing valuable assistance in understanding the CHO cell dataset. We thank Vladimir Porokhin for curating the metabolic model for CHO and generating the synthetic dataset.

**Conflicts of Interest:** The authors declare no conflict of interest. The funders had no role in the design of the study; in the collection, analyses, or interpretation of data; in the writing of the manuscript, or in the decision to publish the results.